\documentstyle[preprint,aps]{revtex}  
\begin{document}
\preprint{UCLA-NT-9703} 
\draft



\title{\bf  Production and decay of the d* dibaryon} 

\author { Chun Wa Wong}

\address{
Department of Physics and Astronomy, University of California, 
Los Angeles, CA 90095-1547}

\date{13 October 1997}

\maketitle

\begin{abstract}
The production of the isoscalar  $J^\pi = 3^+$ didelta dibaryon 
$d^*$ by proton inelastic scattering from deuteron targets is 
described in a double-scattering Glauber approximation. Each 
scattering changes a target nucleon into a $\Delta$ with the help 
of the isovector tensor force transmitted by $\pi$ and $\rho$ 
mesons. The differential cross section constructed from  
empirical Love-Franey nucleon-nucleon $t$-matrices and a simple 
model of $d^*$ shows a maximum of some 10 $\mu$b/sr at 70$^\circ$ 
(c.m.) for 500 MeV protons. The partial width of the decay $d^* \rightarrow NN$ caused by the exchanges of the same mesons is 
found for this simple model of $d^*$ to be about 9 MeV if the 
$d^*$ mass is 2100 MeV. The implications of these results are 
discussed.
\end{abstract}



\narrowtext

\section{ Introduction }
 
The successes of quantum chromodynamics (QCD) between quarks as 
the fundamental theory of strong interactions have led people to 
expect new hadronic states often dominated by exotic 
Fock-space components \cite{Jaf77,Mul78,LISS95}. In nuclear 
physics, one is particularly interested in new nonstrange 
dibaryons (with baryon number $A=2$) of unusually low mass and 
narrow width that might betray their underlying
quark structures. No such dibaryon has been unambiguously
identified experimentally despite years of search 
\cite{LISS95,Com84,Set90}. 

There are two promising dibaryon candidates, one with unusually 
high mass and one with unusually low mass. The high-mass 
candidate is an isospin 1 structure of width not exceeding 80 
MeV first seen experimentally in the helicity difference 
$\Delta \sigma_L(pp)$ of the total $pp$ cross section at an 
energy that corresponds to a dibaryon mass of 2735 MeV \cite{Aue86}. 
It has been seen more recently at the same mass in the $pp$ 
spin correlation parameter $A_{00nn}$ \cite{Bal94}. 

This structure has been interpreted by Lomon and collaborators \cite{Gon87} as a six-quark ``small-bag'' state with the 
nucleon-nucleon ($NN$) quantum number of $^1S_0$. The 
interpretation uses the R-matrix formalism to 
determine if the matching of an internal quark description based 
on QCD to an external nucleon description based on meson-exchange dynamics at a boundary radius separating the two regions could be 
made in a way consistent with the empirical $NN$ phase parameters 
in the neighborhood of the observed structure. (For an assumed 
quark model of internal wavefunctions, the resonance energy can 
be predicted by varying the matching radius $r_0$ until the 
external $NN$ wavefunction vanishes at $r_0$ at precisely 
the same c.m. energy as the internal bag-state energy.) The 
R-matrix analysis also yields a resonance width of about 50 MeV, 
in rough agreement with experiment. This interpretation will 
require confirmation by phase-shift analysis or by 
direct detection via resonance production in nuclear reactions.

In the R-matrix analysis, the experimentally observed mass 
(2735 MeV) has been found \cite{Gon87} to be consistent with the 
bag parameters used in the ``Cloudy-Bag'' model of \cite{Tho81},
provided that the pion cloud is neglected. Furthermore, because of 
its coupling to the external $NN$ channel, the internal bag state is 
not in equilibrium, and therefore the resonance mass is higher than 
the equilibrium bag mass of 2680 MeV \cite{Lom93}. If the external 
pion cloud had been present, as in the actual Cloudy-Bag model \cite{Tho81}, the equilibrium mass (at 2380 MeV) would have been 
much lower \cite{Mul83}. It thus appears that the role of the pion 
cloud external to the bag needs clarification. 

The R-matrix analyses have shown that $NN$ phase parameters are 
also consistent with many other ``high-mass'' nonstrange 
dibaryons \cite{Lom93}, such as those predicted by 
nonrelativistic potential models with pairwise color 
confinement \cite{Won82}. It is obvious that the case for these 
high-mass dibaryons can be significantly strengthened if 
experimental effects are observed for at least another of these dibaryons. Experimental structures seen in $\Delta \sigma_L(pp)$ 
at 2900 MeV \cite{Aue89} and in $\Delta \sigma_L(np)$ at 2630 MeV \cite{Adi96} could be candidates for a $^3P_1$ and a $^3S_1$ 
dibaryon, respectively \cite{Lom97}. 

In addition, suggestions have been made that these nonstrange 
dibaryons might appear at much lower masses instead. 
One with a proposed mass 2065 MeV 
and width $\Gamma_{\pi NN} = 0.5$ MeV might be responsible 
for a narrow structure in the energy dependence of the 
experimental excitation function [at $5^\circ$ in the center of 
mass (c.m.)] of the pionic double charge exchange (DCX) reaction 
$nn(\pi^+,\pi^-)pp$ on nuclear neutrons at $T_\pi = 50$ MeV \cite{Bil92}. This explanation seems to be supported by 
observations of a narrow structure at 2060 MeV with a width 
$< 15$ MeV at ITEP \cite{Vor94} and at CELSIUS \cite{Bro96}. 
The associated dibaryon, usually called $d'$, has the proposed 
quantum numbers $T=0, J^\pi = 0^-$, making it inaccessible from 
$NN$ channels and copnsequently narrow. However, an alternative explanation of the DCX phenomenon that requires no dibaryon has 
also been given \cite{Kag94}.

The dibaryon interpretation finds theoretical support in bag 
models of dibaryon masses where this particular dibaryon 
appears at 2100 MeV\cite{Mul80} or 2000 MeV \cite{Kon87}. 
The theoretical bag state involved is a P-wave excitation in the 
$q^2-q^4$ separation 
with the cluster quantum numbers of $(T,S)_{12} = (0,0)$ and $(T,S)_{3456} = (0,1)$. However, in quark potential models with 
pairwise color confinement, the state appears much higher, at 
around 2700 MeV \cite{Wag95}. The mass can be reduced 
considerably with configuration mixing, but it seems difficult to 
reduce it to below 2400 MeV if one uses quark-quark ($qq$) 
dynamics deduced from single-baryon resonances \cite{Wag95}. 

This paper is concerned with a second nonstrange isoscalar 
dibaryon called $d^*$ which has the quantum numbers 
$J^\pi = 3^+$, is accessible from $np(^3D_3-^3G_3)$ channels, 
and whose baryon-baryon component 
is made up of $\Delta^2$. We shall call this $d^*$ a didelta when 
we want to emphasize this $\Delta^2$ component.

The theoretical $d^*$ mass $m^*$ again covers a wide range: 
It is highest in the small-bag based R-matrix analysis which 
yields a value of 2840 MeV \cite{Lom90}, well above the 
 $\Delta \Delta$ threshold at 2460 MeV. It is lowest in 
the Quark Delocalization and Color Screening (QDCS) Model, where 
it appears at around 2100 MeV \cite{Gold89,Wang92,Gold95}, just 
above the $\pi NN$ threshold at 2020 MeV. Near the lower 
limit of this mass range, the $\pi NN$ phase space is small 
so that the decay of $d^*$ is probably dominated there by the 
$NN$ channel, where the nucleons fall apart in a relative 
D-state. However, the $(\pi)^nNN$ widths could dominate 
as $m^*$ increases.

A search for the $d^*$ dibaryon is interesting for the following 
reasons: In most quark models, the $\Delta-N$ mass difference 
comes from the color-magnetic term of the one-gluon interaction 
between pairs of quarks. The total pairwise color-magnetic 
operator has the same (repulsive) matrix element in $d^*$ as in 
two well separated $\Delta$'s if the former's orbital 
wavefunction is totally symmetric 
in the quark labels \cite{Joh75}. In the Massachusetts Institute of Technology (MIT) bag model \cite{Deg75}, the $d^*$ mass then falls 
below the $\Delta\Delta$ threshold by 120 MeV because the spatial 
integral associated with the interaction is inversely proportional 
to the bag radius $R$, and this radius increases from $\Delta$ to
$d^*$ by virtue of the increasing total kinetic energy \cite{Jaf77,Gold89}. (The radius of the $A$-baryon bag of baryon 
number $A$ is roughly proportional to $A^{1/3}$ in the MIT bag 
model.)

The situation is different if quark confinement comes from a 
pairwise $qq$ interaction that rises to infinity at infinite 
separation. The increasing size of an $A$-baryon containing $3A$ 
quarks causes its confinement energy to increase so much that 
the $d^*$ mass moves 
substantially above the $\Delta\Delta$ threshold instead
\cite{Won82}. This result has been confirmed by \cite{Kal87}.

The QDCS model is able to reduce the $d^*$ mass substantially 
with the help of two additional assumptions: quark delocalization 
(QD) and color screening (CS) \cite{Gold89,Wang92,Gold95}. QD 
takes advantage of the fact that the kinetic energy of a single 
quark state could be reduced if it is partly on the left side and 
partly on the right side of the system. For a Gaussian spatial wavefunction, a maximum reduction of the kinetic energy of about 
19\% appears when there is a 50/50 left/right separation 
with the two wavefunction centers separated by 2.3 oscillator 
lengths, like two peas in a pod. For a dibaryon built up of a 
product of six such delocalized quark wavefunctions, 
about 72\% of the system is in the cluster configurations of 
the type $q^2-q^4$ and $q-q^5$ where the reduction in the 
kinetic energy can be realized. If one reduces the total kinetic 
energy in the MIT bag by the resulting 14\% (72\% of 19\%), one 
gets a $d^*$ mass of about 
2090 MeV instead of the usual 2340 MeV, assuming that the 
interaction and confinement energies retain their spherical forms. 

However, in color confinement models or even in string models, 
this QD reduction of the kinetic energy 
alone cannot overcome the strong increase in pairwise color
confinement energy with increasing $qq$ separations and with 
increasing baryon number $A$. Color screening now comes into play
by assuming that the rising repulsion of the confinement potential 
at large distances \cite{Won82} does not exceed a 
finite upper bound \cite{Gold89,Kal87}. 
Now finally QD could become energetically favorable in $A$-baryons. 
(The model of \cite{Gold95} achieves the same result 
by measuring the confinement energy from the nearer baryon center, 
thus avoiding large $qq$ separations.)

The QD phenomenon could presumably be restricted to the interior 
of a bag. However, if realized between well-separated baryons \cite{Wang92}, the idea has far-reaching implications.
With quark delocalization taking place at all densities, the 
transition to the quark-gluon plasma will be at best a 
second-order phase transition. This seems to imply that if such 
an extreme picture is correct, attempts to search for quark-gluon plasmas might be doomed to failure, given the difficulty of 
detecting unambigous signals from even a first-order phase 
transition. However, a counter-argument is provided by the 
recent observation of unusually strong 
absorption of $J/\psi$ mesons in Pb-Pb collisions \cite{Gon96}. 
One interpretation \cite{Bla96} is that this is a signal 
for the color deconfinement phase transition \cite{Mat86}. 
Consequently, it would be interesting to look for experimental indications for or against the QDCS model. This cannot be done by 
studying nuclear forces, which can already be understood in 
terms of meson exchanges. In contrast, the observation of a low 
$d^*$ mass could be taken as a signal for QD in $A$-baryons. 

In this connection, it is worth noting that a recent R-matrix 
analysis of available $NN$ phase shifts below the dibaryon mass 
of 2240 MeV finds no sign of a $d^*$ resonance in the 
$np(^3D_3-^3G_3)$ channels with a width greater than 1 MeV 
\cite{Lom97}. However, the analysis does not exclude a narrower 
$d^*$ at a mass in between the energies of known phase shifts. 

In any case, progress in our understanding of dibaryons will 
require new experimental inputs. In particular, any new 
information on whether the $d^*$ mass might be high or low 
is likely to have important implications on the dynamics of 
quark confinement in baryons.

The dibaryon $d^*$ could be produced by the inelastic scattering 
of projectiles from nuclear targets. The understanding of past 
failures to find it \cite{Com84,Set90} and the justification for 
future searches in such reactions would require some theoretical 
input concerning its production and decay properties. The main
purpose of this paper is to explore how its production cross 
section in $pd$ inelastic scattering and its partial decay width 
into two nucleons could be calculated using standard techniques 
in nuclear reactions and a simple model of the $d^*$ as a 
didelta object. Some of the issues which must be resolved 
before realistic results can be obtained are briefly discussed.

\section{ The $pd \rightarrow pd^*$ production cross section}

In our treatment of the inelastic production, the deuteron is 
described by an S-state wave function made up of a sum of three Gaussians fitted to the Bonn C S-state wave function \cite{Mac87} renormalized back to 100\%. The d* is taken to be a pure 
$\Delta^2$ Gaussian wave function with an average $\Delta\Delta$ separation of $2r^* = 1.4$ fm \cite{Gold89,Wang92,Gold95}. The 
quark wave function in each baryon is assumed to be the same in 
both $N$ and $\Delta$ \cite{Gold89}. The quarks in 
each baryon are localized to the left or right $\Delta$ in $d^*$, 
with no left-right antisymmetrization or delocalization 
\cite{Gold89} yet included. Later in the paper, we shall estimate qualitatively the effects of delocalization in the calculated 
quantities. 

The excitation of $d(T=0,1^+)$ to $d^*(T=0,3^+)$ requires an 
isoscalar transfer of 2 units of angular momentum. In the usual 
models of nuclear forces containing the exchanges of only 
pseudoscalar, scalar and vector mesons, $d^*$ can only be reached 
in the lowest order by a spin-isospin flip in each of the two 
target nucleons.  Each spin-isospin flip is caused by the 
exchange of an isovector meson such as $\pi$ and $\rho$ between 
the projectile and the target nucleon. It will turn out that 
the unpolarized cross section has important contributions from
that part of the $NN$ $t$-matrix proportional to the operator 
$(\mbox{\boldmath $\sigma$}_1.{\bf q}) (\mbox{\boldmath $\sigma$}_2.
{\bf q})(\mbox{\boldmath $\tau$}_1. \mbox{\boldmath $\tau$}_2)$.

The spin-averaged differential cross section in the c.m. frame has 
the structure

\begin{equation}
{{\rm d}\sigma_{\rm fi} \over {\rm d}\Omega^*} = {p_i^*p_f^*\over \pi}
{{\rm d}\sigma_{\rm fi} \over {\rm d}t} = p_i^*p_f^*
\langle \vert {\cal A}_{\rm fi} \vert ^2 \rangle_{\rm spin}\,, 
\label{eq:dcs}
\end{equation}
where $p_\alpha^*$ is the proton momentum in the reaction c.m. frame 
in the initial or final reaction state $\alpha$. The invariant 
inelastic amplitude 

\begin{equation}
{\cal A}_{\rm fi}({\bf q}) \approx \left({{\rm i}N \over 2\pi}\right)  
\int S_{\rm fi} ({\bf q}, {\bf q}') 
\left [{f_1({\bf q}_1)\over k^*}{f_2({\bf q}_2)\over k^*}\right] 
{\rm d}^2{\bf q}' 
\label{eq:fds}
\end{equation}
is approximated by the Glauber double-scattering contribution 
\cite{Gla70} between quarks in the c.m. frame of the reaction. Each scattering changes a target nucleon into a $\Delta$. The integral 
involves two momentum transfers

\begin{equation}
{\bf q}_1 = \mbox{$1\over 2$}{\bf q} + {\bf q}'\, ,\, 
{\bf q}_2 = \mbox{$1\over 2$}{\bf q} - {\bf q}'\, ,
\end{equation}
and an inelastic formfactor $S_{\rm fi}({\bf q},{\bf q}')$. The 
actual expression is considerably more complicated than 
the symbolic structure shown in Eq.~(\ref {eq:fds}), as we shall 
discuss below. 

The outside factor in Eq.~(\ref {eq:fds}) contains the Glauber 
pair factor $N = N_{\rm T} N_{\rm P}$, where $N_{\rm T} = 9$ for 
the number of distinct quark pairs on the target side, with one 
quark from each of the two target baryons. The effective 
projectile pair number $N_{\rm P}$ is made up of a contribution of 
3 from three projectile quarks each interacting twice with the 
target, and a contribution from three pairs of projectile quarks of

\begin{equation}
3\langle \mbox{$1\over 9$} (\mbox{\boldmath $\sigma$}_1. 
\mbox{\boldmath $\sigma$}_2) (\mbox{\boldmath $\tau$}_1. 
\mbox{\boldmath $\tau$}_2) \rangle = \mbox{$5\over 3$}.
\end{equation}
The formfactors from the projectile quarks are different for 
these two cases, being both different from those for two single 
scatterings from two separate baryons. The difference is described 
by an extra projectile formfactor $S_{\rm P}({\bf q}, {\bf q}')$ 
above and beyond those for two separate elastic $NN$ scatterings. 
Its functional form will be given later.

There are quark-quark operators hidden in the quark-quark 
scattering amplitudes $f_i$ in Eq.~(\ref{eq:fds}). They are 
treated by first expressing $NN$ scattering amplitudes in 
quarks coordinates \cite{Bro75}. After the calculation of 
various operator matrix elements, the scattering amplitudes 
are re-constructed back to $NN$ form.

The momentum $k^*$ is the $NN$ relative momentum in the $NN$ c.m. 
frame. For elastic scattering, or in the high-energy limit
where the inelasticity is negligibly small, it is sufficient to 
use the elastic scattering value 
$k_{\rm el}= (3/4)p_i^*$. We recognize that in 
inelastic scattering at lower energies, the effect of the 
smaller projectile momentum $p_f^*$ in the final state should 
be taken into consideration, in order to describe more accurately 
the  energy dependence of 
both kinematics and dynamics. One possibility is to use the 
geometrical mean momentum $(3/4) \sqrt{p_i^* p_f^*}$, 
but this cannot be correct because it gives the wrong behavior 
at threshold. Of course, the Glauber multiple-diffraction 
formalism is a high-energy approximation that should not be used 
too close to a reaction threshold, but it is conceptually also  
important to ensure that the invariant amplitude $\cal A$ does not 
have a spurious singularity at threshold. The threshold behavior 
is not just a question of kinematics, because the elementary 
amplitude $f_i({\bf q})$ is also involved.

Now the inelastic production of $d^*$ involves at least two 
collisions. At each collision, the energy transfer can 
have a range of values. This suggests that the correct 
description might require an average over some distribution. We 
shall use the simplest realization of this concept, namely the assumption that the $NN$ dynamics in Eq.~(\ref {eq:fds}) should be 
that for the arithmetical average of the proton energies in the 
initial and final states: 

\begin{equation}
T_{\rm av} = T_{\rm lab} - \mbox{$1\over 2$} \Delta m^* ,
\label {eq:tav}
\end{equation}
where $\Delta m^* = m^* - m_d$ is the mass transfer in the 
inelastic scattering. The momentum $k^*$ is the relative momentum 
in the $NN$ c.m. frame when a proton projectile of kinetic energy 
$T_{\rm av}$ is incident on a nucleon target. In this prescription, 
the Glauber double-scattering integral for the invariant amplitude 
does not show a spurious singularity at threshold.

The dynamical factors $(f_i/k^*)$ inside the Glauber 
double-scattering integral is approximately frame invariant, because 
in the usual parametrization using the optical theorem, it depends 
primarily on the total $NN$ cross section $\sigma_{\rm NN}$ if 
nucleons were the elementary objects of the Glauber method.

The reduction of $NN$ amplitudes involving the operator 
$(\mbox{\boldmath $\sigma$}_1.{\bf q}) (\mbox{\boldmath $\sigma$}_2.
{\bf q})(\mbox{\boldmath $\tau$}_1. \mbox{\boldmath $\tau$}_2)$
to quark-quark amplitudes is made in a number of steps:

(a)  We first use the well-known relation \cite{Bro75}
$g_{\rm mqq} = (3/5)g_{\rm mNN}$, where $m$ is $\pi$ 
or $\rho$.

(b)  The $qq$ operators are simplified by using the identities

\begin{equation}
(\mbox{\boldmath $\sigma$}.{\bf q}_1) (\mbox{\boldmath $\sigma$}.
{\bf q}_2) =
{\bf q}_1.{\bf q}_2 + i \mbox{\boldmath $\sigma$}.
({\bf q}_1 \times {\bf q}_2)\, ,
\end{equation}

\begin{eqnarray}
(\mbox{\boldmath $\sigma$}_i.{\bf q}_1) (\mbox{\boldmath $\sigma$}_j.
{\bf q}_2) =
\mbox{$1\over 3$} (\mbox{\boldmath $\sigma$}_i.
\mbox{\boldmath $\sigma$}_j) ({\bf q}_1.{\bf q}_2)  +  
\mbox{$1\over 2$} (\mbox{\boldmath $\sigma$}_i \times
\mbox{\boldmath $\sigma$}_j) . ({\bf q}_1 \times {\bf q}_2) 
+ (\mbox{\boldmath $\sigma$}_i \times
\mbox{\boldmath $\sigma$}_j)^{(2)}.({\bf q}_1 \times {\bf q}_2)^{(2)}.
\label{eq:sigq2}
\end{eqnarray}
Only the tensor-force terms involving 
$(\mbox{\boldmath $\sigma$}_6 \times \mbox{\boldmath $\sigma$}_9)^{(2)} (\mbox{\boldmath $\tau$}_6.\mbox{\boldmath $\tau$}_9)$ 
for two target quarks 6 and 9 contribute to $d^*$ production. 
(Quarks 4-6 are in the first target baryon, while 7-9 are in 
the second.) On the projectile side, we keep only the term that is 
spin-independent. A projectile spin-orbit term can be shown not to contribute to the spin-averaged production cross section under 
rather general circumstances.

The identity (\ref {eq:sigq2}) can also be used to handle central 
spin-isospin dependent interactions by replacing one or more of 
the momentum transfers ${\bf q}_m$ from the tensor interactions by 
projectile quark spin operators $\mbox{\boldmath $\sigma$}_m$ of 
the central interactions. In this way, one can show 
that when both interactions are central, the production of $d^*$ 
induces 2 untis of spin transfer at the projectile, which cannot 
then remain a proton. However, the production $pd \rightarrow pd^*$ 
is possible when one interaction is central and the other is tensor, because in this case the projectile suffers only 1 unit of spin transfer. This mixed tensor-central contribution will not be 
included in the present exploratory study.

(c) In the result reported here, we have neglected all 
quark-exchange terms between target baryons. This permits the 
required operator matrix elements to be calculated in terms of 
the reduced matrix element

\begin{eqnarray}
((N^2) S'T'=10 \parallel (\mbox{\boldmath $\sigma$}_6 \times \mbox{\boldmath $\sigma$}_9)^{(2)} (\mbox{\boldmath $\tau$}_6.\mbox{\boldmath $\tau$}_9) 
\parallel (\Delta^2) 30) 
=  \mbox{$16\over 9$} \sqrt{\mbox{$7\over 2$}}.
\end{eqnarray}
Results with fully antisymmetrized quark wave functions for the 
target will be reported elsewhere when completed. 

On the other hand, the use of $NN$ $t$-matrices ensures that 
all quark exchanges between projectile and target nucleons 
are automatically included.

(d)  The spin-orbit terms in the $NN$ amplitude are neglected, as we 
are not concerned here with polarization phenomena.

The spin-averaged differential production cross section then takes
the form

\begin{eqnarray}
{{\rm d}\sigma_{\rm fi} \over {\rm d}\Omega^*} =  
\left( {1\over 15} \right) 
\sum_{\mu=-2}^2 \vert {\cal F}_\mu({\bf q}) \vert^2\,,
\label{eq:dsig}
\end{eqnarray}
where 

\begin{eqnarray}
{\cal F}_\mu({\bf q}) = \left( {C\over \pi} \right) 
\int {\rm d}^2{\bf q}' 
h({\bf q},{\bf q}') [\mbox{$1\over 4$}({\bf q}\times {\bf q}) - 
({\bf q}'\times {\bf q}')]_{-\mu}^{(2)}\,, 
\label {eq:calF}
\end{eqnarray}
\begin{eqnarray}
C = \left( {N_{\rm P} N_{\rm T}\over 2k^{*2}} \right) \left({3\over 5}\right)^4 
\left({16\over 9}\right) \sqrt{{7\over 2}}\,,
\end{eqnarray}
\begin{eqnarray}
h({\bf q},{\bf q}') = (\mbox{$1\over 4$}q^2 - q'^2)  
\left({m\over 4\pi}\right) 
t_{\rm ivt}({\bf q}_1) t_{\rm ivt}({\bf q}_2) 
S_{\rm P}({\bf q}, {\bf q}') S_{fi}({\bf q}'), 
\label {eq:hqq}
\end{eqnarray}
and $m$ is the nucleon mass. 

We shall use the empirical $NN$ isovector-tensor (IVT) 
$t$-matrix $t_{\rm ivt}$ constructed from $NN$ phase shifts by 
Franey and Love (LF) \cite{Fra85}. The one appearing here is 
actually three times the tensor-force function tabulated by LF, 
i.e. the $NN$ scattering amplitude is here defined as

\begin{eqnarray}
f_{\rm NN}({\bf q}) & = & - \left({m\over 4\pi}\right) 
t_{\rm NN}({\bf q}) \nonumber \\
& = & \left({m\over 4\pi}\right) t_{\rm ivt}(q) 
(\mbox{\boldmath $\sigma$}_1.{\bf q})
(\mbox{\boldmath $\sigma$}_2.{\bf q})
(\mbox{\boldmath $\tau$}_1.\mbox{\boldmath $\tau$}_2) + ...,
\label{eq:tnn}
\end{eqnarray}
where the tensor operator is 1/3 of that contained in the usual 
tensor-force operator used in nuclear physics. 

In the Born approximation, $t_{\rm NN}({\bf q})$ simplifies to 
the Fourier transform $V({\bf q})$ of the $NN$ potential in the 
notation of Ref. \cite{Mac87}. In fact, we shall also use the 
Full Bonn potential \cite{Mac87} in the Born approximation to 
study the separate contributions of $\pi$ and $\rho$ exchanges 
and the dependence of the production cross section on the 
projectile energy.

The additional projectile formfactor needed in Eq.~(\ref {eq:hqq}) is

\begin{eqnarray}
S_{\rm P}({\bf q}, {\bf q}') & = & \exp \left[-{q^2 \over 12b^2} + {q'^2\over 3b^2} \right], \; {\rm for\,single\,quarks;} \nonumber \\
& = & \exp \left[{q^2 \over 24b^2} - {q'^2\over 6b^2} \right], \; {\rm for\,quark\,pairs}.
\end{eqnarray}
Each of these goes with its own effective projectile pair number, 
a minor complication that will not be shown in the formulas.

   On the target side, there are two separate scatterings with two 
baryons. When the inelastic scattering leaves the size of the 
target baryons unchanged, as we have assumed in this paper, the 
baryon formfactors themselves could be reabsorbed back into 
$NN$ scattering amplitudes as they are reconstructed from the 
quark-quark amplitudes. We are then left with just the 
baryon-baryon (or wavefunction) inelastic formfactor

\begin{equation}
S_{\rm fi}({\bf q}) = \left({ 2\beta \beta^* \over \beta^2+\beta^{*2} }\right)^{3/2} 
\exp \left[ -{ q^2 \over 2(\beta^2 + \beta^{*2}) } \right],
\end{equation}
if the deuteron wavefunction is also a single Gaussian, with 
the inverse range $\beta$, while $\beta^* = \sqrt{(3/8)}/r^*$ 
describes the $d^*$ wavefunction of r.m.s radius $r^*$. (This wavefunction radius does not include the contributions from the 
baryon formfactors.) In reality, the Bonn C deuteron S-state 
wavefunction is used after being fitted to the nonorthogonal 
three-term oscillator form

\begin{equation}
\psi_{\rm BonnC}(p) \approx  \sum_{i=1}^3 c_i \psi_i(p),
\end{equation}
where $\psi_i(p)$ is a normalized S-state oscillator wave 
function. The resulting range parameters obtained by minimizing the 
{\it percentage} m.s. deviation are

\begin{equation}
\mbox{\boldmath $\gamma$} =  (\gamma_1, \gamma_2, \gamma_3) = 
0.04467\, (1, 5.04, 21.5)\, {\rm fm}^{-2},
\end{equation}
where $\gamma_i = 2\beta_i^2$. The expansion coefficients, renormalized from the fitted value of 94.34\% back to 100\%, are

\begin{equation}
{\bf c} = (c_1, c_2, c_3) = (0.31491, 0.49716, 0.36926).
\end{equation}

The momentum transfer {\bf q} of the reaction is of course 
calculated with the correct (relativistic) inelastic 
kinematics. However, the other momentum transfers of the 
Glauber approximation are treated in the high-energy limit 
where the inelasticity is negligibly small. As is known, the 
$z$-axis of the Glauber formula is usually not constant in space 
but chosen instead along the bisector of the initial- and 
final-state momenta. This means that the momentum transfer is on 
the equatorial, or $xy$, plane. In fact, all momenta in the 
Glauber double-scattering formula lie on this equatorial plane. In extending the formula to inelastic scattering, we have kept all 
momenta on the equatorial plane even when the energies involved 
are not very high. This means that two of the terms in the sum in 
Eq.~(\ref{eq:dsig}), namely for $\mu = \pm 1$, are zero.

\section{ Results for the production cross section}

The results shown in this section are all calculated in that 
angle-averaged approximation in which $\cos \theta'$ everywhere 
inside the integrand in Eq.~(\ref {eq:fds}) is taken to be 
$1/\sqrt{2}$, $\theta'$ being the angle between 
{\bf q} and ${\bf q'}$. The effect of a full angle integration 
will be discussed near the end of this section. For the 
sake of completeness, we shall give full angular distributions 
even though the Glauber multiple-diffraction approximation is 
known to be reliable only for small angles.

Fig. 1 gives the differential production cross sections at the 
TRIUMF proton energy of 516 MeV. The 1985 Love-Franey (LF) 
$t$-matrix \cite{Fra85} at the lab energy of 515 MeV (on a 
nucleon target) is used for $d^*$ masses of $m^* =$ 2050, 2100 and 
2150 MeV, a range of particular interest for the QDCS model 
mentioned in the Introduction. Calculated with the same $NN$ $t$ 
matrix, these cross sections involve exactly the same invariant production amplitude. They differ only in the outside 
kinematical factors $p_i^*p_f^*$ in Eq.(\ref {eq:dcs}) and in the 
fact that the angular distribution covers a narrower range in 
momentum transfer as $m^*$ increases. For this reason, it is 
often necessary to show results for only 
the smallest $m^* = 2050$ MeV because it has the largest range of momentum transfers.

The cross sections shown are about 7 $\mu$b/sr at the second maximum 
at about $70^\circ$ c.m.. They decrease with increasing $d^*$ 
mass, and should vanish at threshold. The structure at small 
angles, seen very clearly in the $m^*=2050$ MeV result, comes 
from the $\mu = 0$ component of the production tensor operator. 
Its contribution to the differential cross section of 
Eq.~(\ref {eq:dsig}) for $m^*=2050$ MeV is shown separately as a 
short dashed curve in fig.~1. The remaining contributions are from 
the $\mu = \pm 2$ components. Being proportional to $q^4$ for 
small $q^2$, they are responsible for the second maximum. The 
strong $\mu$ dependence means that the angular distributions can 
be expected to be very different when the colliding particles are polarized.

The production mechanism, requiring at least a double scattering, 
is particularly sensitive to the dynamics of the effective $qq$ 
tensor force, here derived from the empirical $NN$ tensor force. 
It is therefore of interest to show, in Fig~2, how strongly the 
production cross section for $m^* = 2050$ MeV at the TRIUMF 
energy increases as the effective nucleon lab energy of the input 
$NN$ $t$-matrix is decreased. 

Since the production cross section as calculated depends only on 
the isovector tensor part of the $NN$ $t$-matrix, the effect seen 
is a direct reflection of the latter's rapid increase with 
decreasing nucleon energy. This behavior is well understood in 
the theory of nuclear forces: The shorter-range part of the tensor 
force due 
to the exchange of $\rho$ mesons is opposite in sign to the 
longer-range part from $\pi$ exchange. As the scattering energy 
decreases, the strong and long-range $\pi$-exchange contribution 
becomes rapidly more dominant. We shall be able to show separately 
the results calculated from each of these two parts of the $NN$ 
tensor force when we use the Full Bonn potential.

The inelasticity involved in $d^*$ production for the $m^*$ range examined here is a very appreciable fraction of the energy 
available in the beam. Given the strong dependence of the 
production cross section on the effective nucleon energy of the 
$NN$ isovector tensor $t$-matrix, it is necessary to choose this 
energy carefully. For reasons discussed in the last section, we 
shall use the average lab energy $T_{\rm av}$ shown in 
Eq.~(\ref {eq:tav}) which is the arithmetical average over the 
initial and final states. 

The resulting cross sections for different $d^*$ masses are shown 
in Fig.~3. These are obtained by interpolating the results 
calculated for the three lab energies $T=$ 515, 425 and 325 MeV tabulated by LF. We see that interpolated results have increased 
by a factor of about 2.0 over the value calculated at the 
incident energy for $m^* = 2100$ MeV. The final cross sections at 
the second maximum are about 13 $\mu$b/sr. This is very large,
but it is perhaps not totally unexpected because the production 
involves the same amplitudes responsible for the resonance 
production of $\Delta$ from nucleon targets.

The use of empirical $t$-matrices takes care of re-scattering 
or wave-distortion effects in the sense of an impulse 
approximation. To determine how important such effects are, we 
use the Full Bonn potential treated in the Born approximation. 
The results obtained with the $NN$ relative momentum $k^*$ 
shown in Eq.~(\ref{eq:fds}) calculated from $T_{\rm av}$ are 
shown in Fig.~4 for different values of $m^*$. 

We see that these production cross sections 
have the same angular behavior as those in Fig.~3 for the LF 
$t$-matrix, but their values are higher by a 
factor of about 2.8. The effect might seem huge, but since 
the calculated cross section is proportional to $g^8$, where $g$ 
is a meson-$NN$ coupling constant, the $t$-matrix reduction of 
the effective $g^2$ is by a very modest factor of 0.77. In other 
words, a strong sensitivity to $NN$ dynamics is unavoidable in such 
a high-order production process. A quantitative calculation of the 
cross section might well require better $t$-matrices constructed 
from more recent $NN$ phase shifts.

The use of the Bonn potential allows the contributions of the 
$\pi$- and $\rho$-exchange potentials to be separated. The 
cross sections for each contribution alone are also shown in 
fig.~4. Note that their total contribution to the cross section is 
not the sum of their separate contributions because amplitudes 
interfere and the production mechanism is double scattering. The 
$\pi$-only result is enormous, but it is effectively controlled by 
the much weaker $\rho$ exchange contribution. Since this 
cancellation is not a well-determined part of $NN$ dynamics, 
Fig.4~ also shows the importance of using the best $NN$ input in 
the calculation.

The Born approximation should improve in accuracy with increasing projectile energy. Fig.~5 shows how the calculated cross section 
for $m^*=$ 2050 MeV decreases with increasing energy, while the 
second maximum moves forward in 
the angular distribution, but not in the momentum transfer. Both features are consistent with the expected energy dependence of 
cross sections. Of course, the accuracy of the Bonn potential at 
these higher energies is somewhat uncertain. In spite of this reservation, Fig.~5 does suggest that the production cross section 
will not decrease sharply with increasing energy.

The production cross section also depends on the $d^*$ 
wavefunction size $r^*$ (half of the average $\Delta\Delta$ 
separation). Fig.~6 gives the results calculated for $r^* = 0.5, 
0.7, 0.9$ fm, covering a realistic range of possible $d^*$ sizes. 
Other parameters used are $m^*=2050$ MeV and an effective 
nucleon energy of 425 MeV for the LF $t$-matrix, very close to 
the recommended average value of 429 MeV. We see that the effect 
is fairly strong especially for smaller $d^*$'s presumably 
because the momentum transfers involved can then be more different. 
We therefore conclude that the calculated cross section 
can be quite sensitive to the short-range components of the wave functions of both $d$ and $d^*$. The short-range components 
neglected in the present calculation include the deuteron D-state, exotic admixtures, and the effects of quark exchange and 
delocalization.

Finally, we address the question of the accuracy of the 
angle-averaged approximation used this section. A selected number 
of angle-integrated cross sections for the TRIUMF energy using the 
Bonn potential with $m^*= 2050$ MeV have been calculated. The 
angle-integrated result at the CM angle 
$\theta^*= 70 (0, 180)^\circ$ is 40.31 (31.06, 6.247) $\mu$b/sr 
when the angle-averaged approximation gives 
38.54 (32.06, 6.252) $\mu$b/sr instead. Thus the angle-averaged approximation seems to have adequate accuracy.  

The present calculations have given only a crude picture of the inelastic production of $d^*$. Their many limitations will be 
discussed in the last section.

\section{ The $d^*\rightarrow NN$ decay width}

The dibaryon $d^*$ cannot decay into two nucleons if its 
constituents do not interact. The simplest interaction that can do 
it is a two-quark interaction containing the spin $(\mbox{\boldmath $\sigma$}_i \times \mbox{\boldmath $\sigma$}_j)^{(2)}$, to bridge 
the gap between initial and final intrinsic spins. If $d^*$ is a didelta, 
even when these deltas are not pointlike, the perturbing operator 
must also contain the isospin operator $\mbox{\boldmath$\tau$}_i . \mbox{\boldmath $\tau$}_j$ in order to change a $\Delta$ into an 
$N$ at each quark vertex. It is clear that isospin-independent interactions cannot do it. Thus all direct gluon 
exchanges do not contribute, no matter how strong they are.

Among the simplest two-quark interactions that can do it are the 
one-meson-exchange potentials carried by $\pi$ and $\rho$. We shall 
show below that their contributions are quite large.

It can be argued that these mesons might not have the presence 
inside $d^*$ that they have in the outside meson cloud, because 
much of the dibaryon interior might be in a different vacuum state, 
the so-called perturbative QCD vacuum, where mesons loose their individuality. Since QCD is flavor-independent, the required
$\mbox{\boldmath $\tau$}_i.\mbox{\boldmath $\tau$}_j$ 
operator could only come from the Heisenberg isospin exchange 
operator 

\begin{equation}
{\cal P}^\tau = \mbox{$1\over 2$}
( 1 + \mbox{\boldmath $\tau$}_i.\mbox{\boldmath $\tau$}_j )
\end{equation}
arising from the Pauli exchange of two quarks. We now show that 
the resulting potential has a functional form similar to that for 
one-meson exchange. In this preliminary study, 
we shall not consider any 
three-quark interactions, including those where two interacting 
quarks involve a noninteracting quark via Pauli exchange, because 
they have more complicated structures.

The exchange two-quark interaction generated by one-gluon exchange 
(OGE) between quarks $i, j$ can be written in the familiar form
\cite{Whe37,Gle83}

\begin{equation}
V_{\rm xqq} = - {\cal P}_{ij} V_{\rm qq} = - {\cal P^\lambda
P^\tau P^\sigma P^{\it x}} V_{\rm qq}, 
\end{equation}
where 

\begin{equation}
{\cal P}^\lambda = \mbox{$1\over 2$} \left( \mbox{$2\over 3$} +
\mbox{\boldmath $\lambda$}_i.\mbox{\boldmath $\lambda$}_j \right)
\end{equation}
is the color exchange operator. The space exchange operator 
${\cal P}^x$ interchanges the spatial labels in the $qq$
final state. For an exchanged gluon of effective mass $\mu$, 
taken here to be 300 MeV \cite{Won96}, the direct $qq$ interaction 
from OGE has the standard one-boson-exchange form \cite{Mac87}

\begin{eqnarray}
 V_{\rm qq}({\bf q}) =  {V_0\over \mu^2 + q^2} 
(\mbox{\boldmath $\lambda$}_i.\mbox{\boldmath $\lambda$}_j) 
\left[ - (\mbox{\boldmath $\sigma$}_i .
\mbox{\boldmath $\sigma$}_j)q^2  +  
(\mbox{\boldmath $\sigma$}_i.{\bf q}) 
(\mbox{\boldmath $\sigma$}_j.{\bf q}) + ... \right] \, .
\end{eqnarray}
The QCD coupling constant appearing in $V_0$ is dependent on the 
gluon mass $\mu$. In addition, it should be an effective 
coupling constant that might include certain higher-order effects. 

Of the terms shown, the second term is a tensor interaction which 
tends to be unobtrusive in baryon spectroscopy, just like tensor 
forces in nuclear spectroscopy. However, the first term is the 
color-magnetic interaction which can be related to the 
$\Delta-N$ mass difference $\Delta m = m_\Delta - m_N = 293$ MeV:

\begin{equation}
V_0 = {\Delta m\over 16 I_B}.
\end{equation}
Here

\begin{equation}
I_{\rm B} = \left\langle {q^2\over q^2 + \mu^2}\delta_i \delta_j
\right\rangle
\end{equation}
is the spatial two-quark matrix element in a baryon (B), taken for simplicity to be the same in both $\Delta$ and $N$. There are 
momentum-conserving $\delta$-functions $\delta_k$ for the two 
quarks $i, j$ at each gluon-quark-quark vertex. 

Is this empirical $qq$ interaction strong or weak? 
This question can be answered by anticipating that even though 
both $V_0$ and the gluon propagator in $V_{\rm qq}$ depend 
significantly on the gluon mass $\mu$, their effects tend to 
cancel when the same $\Delta-N$ mass difference is fitted. The interaction strength $V_0$ turns out to be 0.11 GeV$^{-2}$ when 
$\mu$ is taken to be the $\rho$ meson mass. In contrast, the 
equivalent strength of the one-rho-exchange (ORE) potential in the 
Full Bonn potential is

\begin{equation}
V_0 = \left( {3\over 5} \right)^2 {\pi\over m^2} {g_\rho^2\over 4\pi} 
\left(1 + {f_\rho\over g_\rho}\right)^2 = 54 \; {\rm GeV}^{-2},
\end{equation}
where the factor $(3/5)^2$ comes from the reduction from $NN$ to 
$qq$ operators. This is about 500 times stronger than the $qq$ 
interaction strength. Although the $\rho NN$ coupling constant is 
not well determined in the $NN$ interaction, it is clear that the 
ORE potential appropriate to the baryon exterior is 
some two orders of magnitude stronger than 
the effective $qq$ interaction present in the perturbative vacuum 
of the baryon interior. This shows that the OGE contribution to 
the partial decay width is negligibly small compared to the 
meson-exchange contributions. 

It is nevertheless interesting to complete the derivation of the exchange $qq$ interaction and to determine how the 
space-exchange operator ${\cal P}^x$ further affects 
the final result. We therefore go on by noting that the 
term in $V_{\rm xqq}$ containing the $d^* \rightarrow NN$ decay 
operator can be isolated by using the expansion

\begin{eqnarray}
- {\cal P^\lambda P^\tau P^\sigma}  
(\mbox{\boldmath $\lambda$}_i.\mbox{\boldmath $\lambda$}_j) 
 (\mbox{\boldmath $\sigma$}_i.{\bf q}) 
(\mbox{\boldmath $\sigma$}_j.{\bf q}) = 
- \mbox{$8\over 9$}
(\mbox{\boldmath $\sigma$}_i.{\bf q}) 
(\mbox{\boldmath $\sigma$}_j.{\bf q})
(\mbox{\boldmath $\tau$}_i.\mbox{\boldmath $\tau$}_j) + ... \, ,
\end{eqnarray}
obtained with the help of the identities

\begin{equation}
(\mbox{\boldmath $\lambda$}_i.\mbox{\boldmath $\lambda$}_j)^2
=  \mbox{$32\over 9$} - \mbox{$4\over 3$}
(\mbox{\boldmath $\lambda$}_i.\mbox{\boldmath $\lambda$}_j)
\end{equation}
and

\begin{eqnarray}
(\mbox{\boldmath $\sigma$}_i.\mbox{\boldmath $\sigma$}_j) 
(\mbox{\boldmath $\sigma$}_i.{\bf q}) 
(\mbox{\boldmath $\sigma$}_j.{\bf q}) = 
(\mbox{\boldmath $\sigma$}_i.{\bf q}) 
(\mbox{\boldmath $\sigma$}_j.{\bf q}) + 
 (1 - \mbox{\boldmath $\sigma$}_i.\mbox{\boldmath $\sigma$}_j) q^2.
\end{eqnarray}
The final result is

\begin{eqnarray}
V_{\rm xqq}({\bf q}) = - {\cal P}^x \left( {8\over 9} \right)
(\mbox{\boldmath $\sigma$}_i.{\bf q}) 
(\mbox{\boldmath $\sigma$}_j.{\bf q}) 
(\mbox{\boldmath $\tau$}_i.\mbox{\boldmath $\tau$}_j) 
{V_0\over \mu^2 + q^2} 
+ ... \, , 
\label{eq:vxqq}
\end{eqnarray}
showing only the term which can turn colorless $\Delta$'s into 
colorless $N$'s. This term too has a strength characterized by 
$V_0$. This is the only term that can contribute, in the lowest 
order, to the decay of the $d^*$ treated as a didelta.

In more realistic models of $d^*$, colored $\Delta$'s appear in 
the so-called ``hidden-color'' components \cite{Har81}. These 
colored objects can decay into $NN$ by both direct and exchange 
OGE potentials via terms proportional to the color operator
$\mbox{\boldmath $\lambda$}_i.\mbox{\boldmath $\lambda$}_j$. 
However, all such contributions are necessarily based on the 
effective $qq$ interaction $V_{\rm qq}$ whose strength $V_0$ is 
two orders of magnitude weaker than meson-exchange interactions. 

We are now in a position to calculate the decay width. Its
spin-averaged value in first-order perturbation theory 
\cite{Sak67} has the structure 

\begin{eqnarray}
\Gamma(m^*) =  2\pi p^*\mu_f^*
\int {\rm d}^2\Omega_{\bf p^*} 
\vert N_{\rm qq}\langle (N^2){\bf p}^*\vert V \vert(\Delta^2)d^* \rangle \vert^2_{spin}\,, 
\label{eq:width}
\end{eqnarray}
where ${\bf p}^*$ is a nucleon momentum in the center-of-mass 
frame and $\mu_f^*$ ($=m^*/4$) is the relativistic reduced mass 
in the final state calculated with dynamical masses or total 
energies. $N_{\rm qq} = 9$ pairs of $qq$ interaction 
$V$ contribute to the decay. The momentum state 
$\vert {\bf p}^* \rangle$ has the normalization:

\begin{equation}
\langle {\bf r}\vert {\bf p}^* \rangle = 
{1\over (2\pi)^{3/2}} e^{i {\bf p}^*.{\bf r}}.
\end{equation}

After some algebra, the decay width can be written in the final form 

\begin{eqnarray}
\Gamma(m^*) =  { p^*\mu_f^*\over 7}  \left[ N_{\rm qq} {16\over 9} 
\sqrt{{7\over 2}}  \right]^2 {16\over 15} 
{1\over (2 \pi)^2 \pi^{3/2}}
\left( \beta^{*7}\over \kappa^6 \right) 
F_{\rm model},
\end{eqnarray}
where the model-dependent factor $F_{\rm model}$ is

\begin{equation}
F_{\rm mx} = \left( \mbox{$3\over 5$} \right)^2 
I_{\rm ivt}(\kappa, 1, 1, 1)
\end{equation}
for the meson-exchange (MX) model, and is

\begin{eqnarray}
F_{\rm xqq} = \left( {8\over 9} \right)
\left( {9\beta^{*2}\over 5\beta^{*2} + 6b^2} \right)^{3/4}
I_{\rm xqq}(\kappa, D_1, D_2, D_3)
\end{eqnarray}
for the exchange-$qq$ (XQQ) model. Here

\begin{eqnarray}
I_{\rm model}(\kappa, D_1, D_2, D_3) = 
{\rm e}^{-D_1\kappa^2/2} 
 \int {\rm e}^{-D_2Q^2/2}[j_2(iD_3\kappa Q)(\kappa Q)^3] 
t_{\rm model}(\beta^*Q)Q {\rm d}Q\,,
\end{eqnarray}
$\kappa = k^*/\beta^*$ is the dimensionless nucleon momentum in 
the c.m. frame, and $Q = q/\beta^*$ is the dimensionless momentum transfer. The elastic baryon form factor at each quark vertex is 
already contained in the $NN$ interaction $t_{\rm ivt}$ of 
Eq.~(\ref {eq:tnn}). Furthermore, the $t$-matrix contains $NN$ rescattering contributions to all orders of the $NN$ interaction.

For the XQQ model, for which $t_{\rm xqq}$ 
in this equation refers to the function 
$V_0/(\mu^2 + q^2)$ in Eq.~(\ref{eq:vxqq}), the exchange 
baryon formfactor appears explicitly in $F_{\rm xqq}$. The 
latter depends on the dimensionless parameters $D_i$ 
which are functions of the baryon size parameters $\beta^*$ 
and $b = 1/r_p$, where $r_p = 0.6$ fm is the proton radius. 
For $r^* = 0.7$ fm, they have the numerical values of

\begin{eqnarray}
D_1 = 0.214;\; D_2 = 0.398;\; D_3 = 0.153.
\end{eqnarray}

\section{ Results for the partial decay width}

The partial decay width for $d^* \rightarrow NN$ can now be 
calculated numerically for the empirical Love-Franey 
$NN$ $t$-matrix, the Full Bonn potential treated in the Born approximation, and the OGE $qq$ interaction with Pauli exchange. 

The empirical LF $t$-matrix used is evaluated at the equivalent 
nucleon lab energy 

\begin{equation}
T_{\rm lab} = {m^{*2}\over 2m} - 2m
\end{equation}
of the final $NN$ state. The resulting decay widths $\Gamma$, 
obtained by interpolating from the values calculated at the 
three nearest energies tabulated by LF, are shown in Fig.~7 as 
functions of the $d^*$ mass $m^*$ for different $d^*$ wavefunction 
radii $r^*$. The dependence on these parameters is significant, 
but not strong, and is of a somewhat complex character.

The origin of some of the complexities in the behavior of $\Gamma$ 
can be seen in Fig.~8, which gives the corresponding results for
$r^* = 0.7$ fm covering a wider range of dibaryon mass calculated 
with the Full Bonn potential treated in 
the Born approximation. The result is typically 2--3 times larger 
than those for the LF $t$-matrix in the mass range (2050-2130 MeV) 
of interest in the QDCS model, in qualitative agreement with the 
corresponding behavior in the production cross section. Note 
however that the leading decay process studied here is first-order 
in the interaction, but the leading production process is second 
order. 

The ``$\pi$ only'' contribution (shown as a dash-dot curve) 
is again much larger, here by another factor of three in this 
mass range. Even the ``$\rho$ only'' result (given by the short 
dashed curve) is large, becoming in fact arger than even the 
``$\pi$ only'' contribution after 2690 MeV. 
This behavior is consistent with the fact that the decay 
involves interactions at shorter distances and higher energies
than the production at TRIUMF energies. This means that the 
results for $\Gamma$ is particularly sensitive to the 
cancellation between the $\pi$ and $\rho$ exchange interactions 
in our simple leading-order model. In fact, the vanishing of the 
calculated decay width due to the complete destructive 
interference between the $\pi$ and $\rho$ contributions
can be seen in the figure at 2690 MeV. 

For the XQQ interaction appropriate to the perturbative vacuum of 
the baryon interior, we find a result of $\Gamma = 77$ eV at 
$r^* = 0.7$ fm, $m^* = 2100$ MeV and a gluon mass of $\mu = 300$
MeV if $V_{\rm xqq}$ interaction of eq.~(\ref {eq:vxqq}) is treated 
as a direct $qq$ interaction. (This is done by dropping the 
space-exchange operator ${\cal P}^x$ and adding baryon elastic formfactors to the expression for $\Gamma$.) Compared to the result 
of 22 MeV for ``$\rho$ only'', this is a factor of $3 \times 10^5$ smaller, in good agreement with the estimate based on the 
interaction strengths.

On restoring the space-exchange operator ${\cal P}^x$, we find that 
the width drops drastically by another factor of $3 \times 10^6$ to 
only $2.1 \times 10^{-5}$ eV. The result is very sensitive to $r^*$, being 250 times larger for the smaller $d^*$ radius of $r^* = 0.5 $ 
fm. One reason why the result is so small is that the orbital angular momentum in the relative baryon-baryon coordinate is $L=2$, where the centrifugal potential inhibits the quark exchange. If we had used 
$L=0$ instead, the result would have been $2.3 \times 10^{-3}$ eV, 
as compared to an $L=0$ width of 190 eV when ${\cal P}^x$ is 
purposely dropped. We conclude from this that in the absence of significant short-range components in the wave function, 
quark-exchange effects are negligible. 
 
Up to this point, we have done the calculation as if the 
interaction were 100\% by meson exchanges via the meson cloud in the baryon exterior, or 100\% by gluon exchange in the perturbative 
vacuum of the baryon interior. In reality, there is always some ``external'' contribution even when all the six quarks of $d^*$ are
in the same cluster. Since the external contribution is so much 
greater than the internal contribution, some of it will always 
survive to give a nonnegligible decay width. In the absence of a detailed model describing the precise proportion between the two contributions, we shall discount our calculated meson-exchange contribution by 50\% as a guess of what a more realistic width 
should be.

The interior correction for gluon exchange discussed here does not 
apply to the lowest-order production calculated in Sec.~3: The two mesons whose exchanges between projectile and target nucleons are responsible for the inelastic production of $d^*$ are both external virtual mesons.

Is the meson-exchange interactions used here too strong? This can 
be answered partially by using the same method to calculate the 
decay width 
of that archtypical baryon decay $\Delta \rightarrow \pi N$. Using 
the $\pi NN$ coupling constant of $f_{\pi {\rm NN}}^2/4\pi = 0.078$ given by the Full Bonn potential and reducing to a $\pi qq$ 
coupling constant, we find a result of only 70 MeV, much less than 
the experimental value of 120 MeV. This result is in agreement 
with past calculations of this width \cite{Bro75}. One might be 
tempted to increase the calculated result by a factor of 
$(120/70)^2 \approx 3$, but this is not advisable because a real 
$\pi$ on the energy shell is emitted in the $\Delta$ decay, 
whereas the $d^* \rightarrow NN$ decay involves virtual mesons off 
the energy shell \cite{Sug69}. The off-shell coupling constants 
that appear are more appropriately determined from nuclear 
forces, at least in principle. If we had been calculating the 
$d^* \rightarrow \pi NN$ decay by the same method, we would be 
justified to increase the calculated result by a factor 
$120/70 \approx 1.7$ for that vertex emitting the real pion.

\section { Dependence on quark models }

The calculated $d^*$ production cross section and decay width 
can be expected to depend sensitively on the quark wavefunction 
of the $d^*$, perhaps even more so than its theoretical mass. 
We shall consider qualitatively some of the issues involved.

One of these issues is the possibility of quark delocalization, 
which refers to the idea that under certain circumstances a quark 
may find it energetically favorable to be partly on the left side 
and partly on the right side of a dibaryon. The QDCS model 
\cite{Gold89} actually describes each quark in the $d^*$ 
wavefunction as 50/50 left or right. The six-quark $d^*$ 
wavefunction then has the structure

\begin{eqnarray}
(L+R)^6 & = & L^6 + 6 L^5 R + 15 L^4 R^2 + 20 L^3 R^3 
+ 15 L^2 R^4 + 6 L R^5 + R^6 \nonumber \\
& \rightarrow&  2 (L^6\,{\rm  or}\, R^6) + 12 (L^5 R \,{\rm or}\, L R^5) 
+ 30 (L^4 R^2 \,{\rm or}\, L^2 R^4) + 20 (L^3 R^3).
\end{eqnarray}
After the projection of relative S-states between the clusters and a correction for the position of the center of mass, the wavefunction simplifies to the form shown after the right arrow. Its components 
fall roughtly into two groups: There is a group of normal (``n'') clusters of $q^{3m}$ configurations made up of $L^3 R^3, L^6$ and 
$R^6$ with no delocalized quark. They have the probability of 

\begin{equation}
P_{\rm n} = (2^2 + 20^2)/1448 \approx 0.28.
\end{equation}
The remaining group of components $L^5 R, L R^5, L^4 R^2,$ and 
$L^2 R^4$ has one delocalized quark (``dq'') away from a normal 
$q^{3m}$ configuration, and the remaining probability of 
$P_{\rm dq} = 1 - P_{\rm n} \approx 0.72.$

For the normal group, the projection of S-states makes the 
wavefunction spherical symmetric in the relative baryon-baryon coordinate, and very similar to the Gaussian wavefunction of our 
didelta model. In fact, the maximum overlap between the two 
wavefunctions is close to 100\%. There is to be sure some depression 
of the two-center relative wavefunction near the origin of the 
relative coordinate, but the effect is quite unimportant in the wavefunction overlap. The behavior of the short-distance 
wavefunction is probably much more important in the production and 
decay processes considered in this paper, but it is likely that the short-distance wavefunction is not very good in both models. Furthermore, the $q^6$ component is entirely absent in the didelta 
model and is probably too weak in the QDCS model. 

Though subject to these additional uncertainties, our counting 
suggests that this normal group will contribute essentially the 
full amount, i.e. about 28\% of that calculated in our model in 
both decay and production.

For the abnormal components with one ``wrong-way'' quark, the contribution could be very different, especially if there is 
special coherence between the normal and abnormal amplitudes. We 
are not in a position to estimate such coherent contributions 
because it would require a specific model. As far as its 
incoherent contribution is concerned, the worse that can happen 
is that it will vanish. This must be a rather extreme situation, 
because three of the nine pairs of interacting quarks involve the ``wrong-way'' quark, and the interaction could scatter it back to 
form a normal cluster structure. In the remaining six pairs, the ``wrong-way'' quark is a spectator, which requires a wavefunction overlap to get back to normal. There is thus a reduction in the calculated decay or production amplitude of the order of $1/2$ or 
1/e for the spectator contribution, more if the clusters are 
farther apart. We then end up with an estimate for the decay width 
or production cross section of the order of 7--20\% from these 
abnormal components.  

Thus very crudely, we expect the delocalization to reduce the 
calculated decay width or production cross section by a factor of
1/2--1/3.

Another model-dependent issue is the contribution of hidden-color 
(HC) configurations. Our didelta model used without quark 
exchanges between the two baryon clusters contains no HC component. 
In contrast, most quark models of the $d^*$ contains 80\% HC 
components where the first three quarks are in a color-octet 
state. These HC components are expected to contribute less, 
perhaps significantly less, than the baryon-baryon components. A calculation of their contributions using the method of this paper 
is now underway. For the time being, we shall allow for some contributions from the HC components by reducing the 
baryon-baryon results by a factor of 1/2. (The reduction factor is 
1/5 when the HC components contribute nothing.)

The final educated guesses for $d^*$ decay and production for the 
quark-delocalization model used with the Love-Franey NN $t$-matrix 
at $m^* = 2.1$ GeV and $r^* = 0.7$ fm are as follows: The decay 
width is decreased from 9 MeV to about 1 MeV when an ``interior'' correction of 1/2 and an octet-octet reduction factor of 1/2 are
also included. The production cross section at TRIUMF energy is 
reduced from 13 to about 2 $\mu$b/sr at the second maximum.

\section { Discussions and conclusions }

The leading-order processes studied here suggest that the inelastic production cross section of $d^*$ might be in the $\mu$b/sr range, 
while its decay width into two nucleons might be in MeV's. 
Love-Franey empirical $NN$ $t$-matrices are used to include all 
$NN$ re-scattering effects to all orders in the decay and in an 
impulse approximation for the production. 
Other aspects of the reported calculations are not sufficiently realistic because of approximations made in the hadron 
wavefunctions and in the treatment of the reaction mechanisms. 
It is worthwhile to enumerate the most serious of these problems.

The Glauber multiple-diffraction model used in the calculation of 
the production cross section might be quite good for elastic 
scattering at small angles. Its validity for large inelasticities 
and at large angles is unknown. It is necessary to 
correct for effects neglected by the Glauber model, especially at 
large angles \cite{Won75}. However, there is probably no point in 
doing this unless one can also include higher-order production processes. These are the usual difficulties connected with the 
calculation of strong-interaction cross sections, and as usual, we 
see no simple solution.

The Gaussian model of $d^*$ used here is very crude. Depending 
on the model, we need to add delocalization and short-distance 
refinements. The S-state wave function used for the deuteron 
target is also inadequate. It is not difficult to put in the 
D-state $NN$ component. Other short-range components such as 
hidden-color configurations can readily be treated too. The 
difficulty lies instead in the lack of knowledge on how strong 
these components are. Studies of the effects of short-distance wavefunctions require sustained efforts. 

In the calculation of the partial $d^* \rightarrow NN$ decay width, 
it is necessary to account more carefully for the role played by  external meson exchanges versus internal gluon exchanges. It is 
obvious that this too cannot be done on a quantitative basis 
without a more realistic model of the $d^*$ wavefunction. It will 
be necessary to include higher-order processes not yet included by 
using $NN$ $t$-matirces.

In addition, the decay width could be dominated by the 
$(\pi)^nNN$ branches for sufficiently large $d^*$ mass. It is 
necessary to understand these partial decay widths at least qualitatively.

The theoretical picture concerning the dibaryon $d^*$ at the 
present time seems to be as follows: Its calculated mass has 
been in the range 2050\cite{Gold89}--2840\cite{Gon87} MeV. Its 
inelastic production cross section could be significant, i.e. in 
the $\mu$b/sr range. Its partial decay width into two nucleons 
is probably in MeV's for the low-mass candidate. Because 
its calculated mass is so sensitive to certain assumptions 
concerning quark dynamics in hadrons, any positive or negative experimental information on its presence in a certain mass range 
has interesting implications.

What is the experimental situation concerning isoscalar 
dibaryons? A dibaryon search was made at Saturne by measuring 
the spectra for missing masses between 1.9 and 2.35 GeV using 
the $dd \to dX$ reaction for 2.29, 2.00 and 1.65 GeV deuteron 
beams. An upper limit of 30 nb/GeV$^2$ was found for the 
invariant production cross section of a dibaryon if its width is 
less than several tens of MeV \cite{Com84}. This result is for 
the missing mass of 2.16 GeV and for a 2.00 GeV deuteron beam 
with deuterons detected at $27^\circ$ (lab), or $69.3^\circ$ (c.m.). 
It corresponds to a c.m. differential production cross section 
at this angle of only 15 nb/sr in the $dd$ reaction. 
It is not easy to extract a $pd$ bound from this result 
partly because of the presence of an elastic formfactor for the 
intact deuteron \cite{Yen97}, which causes a large reduction in 
the $dd$ production amplitudes relative to the $pd$ amplitude for production from single nucleons in the intact deuteron.
An additional complication is that for the $dd$ reaction, the $d^*$ 
can also be produced by another double-scattering process that 
involves both nucleons of the intact $d$. Its contribution can be expected to be similar in structure, but probably reduced in value, 
when compared to that of production from a single 
nucleon in the $pd$ reaction. This process must also be included 
in the interpretation of $dd$ cross sections. This means that any extraction of a $pd$ bound from the $dd$ bound will depend 
on a model-dependent theoretical analysis, and cannot be a pure experimental bound. 

To my best knowledge, the only direct experimental upper bound 
for resonance production in the $pd$ reaction is an unpublished 
LAMPF experiment based on the $d({\vec p},p)X^+$ reaction at 
$T_p = 798$ MeV and $15.1^\circ$ (lab).  The results are in the 
range of 1--4 $\mu$b/sr/MeV \cite{Set90} dependent on the missing 
mass in the missing-mass range of 1865--2200 MeV. They are of the 
same order as the very rough theoretical estimates of $d^*$ 
production given in this paper. Hence no definite conclusion can 
be drawn from a comparison between them.

It thus appears that the present theoretical picture is still 
very unrealistic and incomplete. Much additional work is 
needed, especially on the partial decay widths 
in pionic channels for which there is at present very little 
quantitative information. However, the question of dibaryons is 
ultimately an experimental question. A new dibaryon search with 
a sensitivity much greater than the known LAMPF bound will be 
needed to advance our understanding of the situation.

\acknowledgements

I would like to thank Fan Wang, Stan Yen, Terry Goldman, Earle 
Lomon, Gary Love, Mike Franey and Mahmood Heyrat for many 
stimulating discussions and correspondence.

\begin{figure}
\caption{Center-of-mass differential cross section for $pd\to pd^*$ 
at the proton lab energy of 516 MeV for different $d^*$ masses $m^*$
using the 1985 Love-Franey $t$-matrix at 515 MeV. The $m^* = 2050$ 
MeV result from only the $\mu = 0$ term of Eq.~(9) is also shown.}
\label{fig1}
\end{figure}

\begin{figure}
\caption{ Differential cross sections for $pd\rightarrow pd^*$ at 
516 MeV for different Love-Franey $t$-matrix energies using
$m^*= 2050$ MeV.}
\label{fig2}
\end{figure}

\begin{figure}
\caption{ Differential cross sections for $pd\rightarrow pd^*$ at 
516 MeV for different $m^*$ masses using the Love-Franey $t$-matrix 
at the  energy averaged over the initial and final states.}
\label{fig3}
\end{figure}

\begin{figure}
\caption{ Differential cross sections for $pd\rightarrow pd^*$ at 
516 MeV for different $m^*$ in the notation of Fig.~3 using the 
Full Bonn potential in the Born approximation with the $NN$ 
relative momentum $k^*$ calculated at an average energy 
$T_{\rm av}$. The $m^*=2050$ MeV results for $\pi$-exchange only 
and for $\rho$ exchange only are also given.}
\label{fig4}
\end{figure}

\begin{figure}
\caption{ Differential cross sections for $pd\rightarrow pd^*$ 
for different projectile energies using $m^*=2050$ MeV, the 
Full Bonn potential in the Born approximation, and an average 
nucleon energy $T_{\rm av}$.}
\label{fig5}
\end{figure}

\begin{figure}
\caption{ Differential cross sections for $pd\rightarrow pd^*$ 
at 516 MeV for different $d^*$ wavefunction radii $r^*$ for 
$m^*=2050$ MeV using the Love-Franey $t$-matrix at 425 MeV.}
\label{fig6}
\end{figure}

\begin{figure}
\caption{ Decay width for $d^*\rightarrow NN$ as a function of 
the $d^*$ masss $m^*$ for different $d^*$ wavefunction radii 
$r^*$ using the Love-Franey $t$-matrix at the final-state energy.}
\label{fig7}
\end{figure}

\begin{figure}
\caption{ Decay width for $d^*\rightarrow NN$ as a function of 
the $d^*$ masss $m^*$ for $r^* = 0.7$ fm using the Full Born 
potential in the Born approximation. The result for the 
Love-Franey $t$-matrix is also shown as a solid curve.}
\label{fig8}
\end{figure}


\begin{references}

\bibitem{Jaf77}
R.L. Jaffe, Phys. Rev. Lett. {\bf 38}, 195 (1977).

\bibitem{Mul78}
P.J.G. Mulders, A.T.M. Aerts and J.J. de Swat, Phys. Rev. D {\bf 17}, 260 (1978).

\bibitem{LISS95}
LISS (Light Ion Spin Synchrotron) White Paper, January, 1995, Indiana University Cyclotron Facility at http://www.iucf.indiana.edu/Publications/LISS.html

\bibitem{Com84}
M.P. Combes {\it et al.}, Nucl. Phys. {\bf A431}, 703 (1984).

\bibitem{Set90}
K.K. Seth {\it et al.}, contributed paper to PANIC-90 Particles and Nuclei Conference (Boston, 1990, unpublished). 

\bibitem{Aue86}
I.P. Auer {\it et al.}, Phys. Rev. D {\bf 34}, 2581 (1986).

\bibitem{Bal94}
J. Ball {\it et al.}, Phys. Lett. B {\bf 320}, 206 (1994).

\bibitem{Gon87}
P. Gonzalez, P. LaFrance and E.L. Lomon, Phys. Rev. D {\bf 35}, 2143 (1987). 

\bibitem{Tho81}
A.W. Thomas, S. Th\'eberge and G.A. Miller, Phys. Rev. D {\bf 24}, 216 (1981).

\bibitem{Lom93}
E.L. Lomon, in Proc. Workshop on future directions in particle and nuclear physics at multi-GeV hadron beam facilities, BNL-52389 (BNL, 1993), pp. 406-411. 

\bibitem{Mul83}
P.J.G. Mulders and A.W. Thomas, J. Phys. G {\bf 9} (1983) 1159;
K. Saito, Prog. Theor. Phys. {\bf 72}, 674 (1984).

\bibitem{Won82}
C.W. Wong, Prog. Part. Nucl. Phys. {\bf 8}, 223 (1982). 

\bibitem{Aue89}
I.P. Auer {\it et al.}, Phys. Rev. Lett. {\bf 62}, 2649 (1989).

\bibitem{Adi96}
B.P. Adisevich {\it et al.} Z. Phys. C {\bf 71}, 65 (1996).

\bibitem{Lom97}
E.L. Lomon, unpublished 1997 report,  nucl-th/9710006 (MIT-CTP-2680).

\bibitem{Bil92}
R. Bilger {\it et al.}, Z. Phys. {\bf A343}, 491 (1992); 
R. Bilger, H.A. Clement and M.G. Schepkin, Phys. Rev. Lett. {\bf 71},
42 (1993); {\bf 72}, 2972 (1994).

\bibitem{Vor94}
L.S. Vorobev {\it et al.}, JETP Lett. {\bf 59}, 77 (1994).

\bibitem{Bro96}
W. Brodowski {\it et al.}, Z. Phys. {\bf A355}, 5 (1996).

\bibitem{Kag94}
M.A. Kagarlis and M.B. Johnson, Phys. Rev. Lett. {\bf 73}, 38 (1994).

\bibitem{Mul80}
P.J. Mulders, A.T. Aerts and J.J. de Swart, Phys. Rev. D {\bf 21}, 
2653 (1980).

\bibitem{Kon87}
L.A. Kondratyuk, B.V. Artemyanov and M.G. Schepkin, Sov. J. Nucl. Phys. {\bf 45}, 776 (1987).

\bibitem{Wag95}
G. Wagner, L.Ya. Glozman, A.J. Buchmann and A. Faessler, Nucl. Phys. {\bf A594}, 263 (1995).

\bibitem{Lom90}
E.L. Lomon, J. Phys. 58 Supp. Colloque C6, 363 (1990).

\bibitem{Gold89}
T. Goldman, K. Maltman, G.T. Stephenson, Jr., K.E. Schmidt and F. Wang,  Phys. Rev. C {\bf 39}, 1889 (1989); F. Wang, J.L. Ping, G.H. Wu, L.J. Teng and T. Goldman, Phys. Rev. C {\bf 51}, 3411 (1995).

\bibitem{Wang92}
F. Wang, G.H. Wu, L.J. Teng and T. Goldman, Phys. Rev. Lett. {\bf 69},
2901 (1992).

\bibitem{Gold95}
T. Goldman, K. Maltman, G.T. Stephenson, Jr., J.L. Ping and F. Wang,  Systematic Theoretical Search for Dibaryons in a Relativistic Model, Los Alamos Preprint LA-UR-95-2609.

\bibitem{Joh75}
K. Johnson, Acta Phys. Pol. {\bf B6}, 865 (1975).

\bibitem{Deg75}
T.A. DeGrand {\it et al.}, Phys. Rev. D {\bf 12}, 2060 (1975). 

\bibitem{Kal87}
Yu.S. Kalashnikova, I.M. Narodetskii and Yu.A. Simonov, Yad. Fiz 
{\bf 46}, 1181 (1987) [Sov. J. Nucl. Phys. {\bf 46}, 689 (1987)]. 

\bibitem{Gon96}
M. Gonin [NA50], Nucl. Phys. A {\bf 610}, 404c (1996);
C. Lourenco [NA50], Nucl. Phys. A {\bf 610}, 552c (1996); 

\bibitem{Bla96}
J.-P. Blaizot and J.-Y. Ollitrault, Phys. Rev. Lett. {\bf 77}, 
1703 (1996). 

\bibitem{Mat86}
T. Matsui and H. Satz, Phys. Lett. {\bf 178B}, 416 (1986). 

\bibitem{Mac87}
R. Machleidt, K. Holinde and Ch. Elster, Phys. Rep. {\bf 149}, 1 
(1987).

\bibitem{Gla70}
R.J. Glauber, in {\it High Energy Physics and Nuclear Structure}, 
edited by S. Devons (Plenum, New York, 1970) p.207.

\bibitem{Bro75}
G.E. Brown and W. Weise, Phys. Rep. {\bf 22C}, 279 (1975).

\bibitem{Fra85}
M.A. Franey and W.G. Love, Phys. Rev. C {\bf 31}, 488 (1985).

\bibitem{Whe37}
J.A. Wheeler, Phys. Rev. {\bf 52} 1083, 1107 (1937).

\bibitem{Gle83}
N.K. Glendenning, Direct nuclear reactions (Academic Press, New York, 1983), pp. 194-195.

\bibitem{Won96}
C.W. Wong, Phys. Rev. D {\bf 54}, R4199 (1996).

\bibitem{Har81}
M. Harvey, Nucl. Phys. A {\bf 352}, 301 (1981).

\bibitem{Sak67}
J.J Sakurai, Advanced quantum mechanics (Addison-Wesley, Reading, MA, 1967), pp. 171-172.

\bibitem{Sug69}
H. Sugawara and F. von Hippel, Phys. Rev. {\bf 185}, 2046 (1969); 

\bibitem{Won75}
C.W. Wong and S.K. Young, Phys. Rev. C {\bf 12}, 1301 (1975); 
S.K. Young and C.W. Wong, Phys. Rev. C {\bf 15}, 2146 (1977).

\bibitem{Yen97}
S. Yen, private communication.

\end{references}
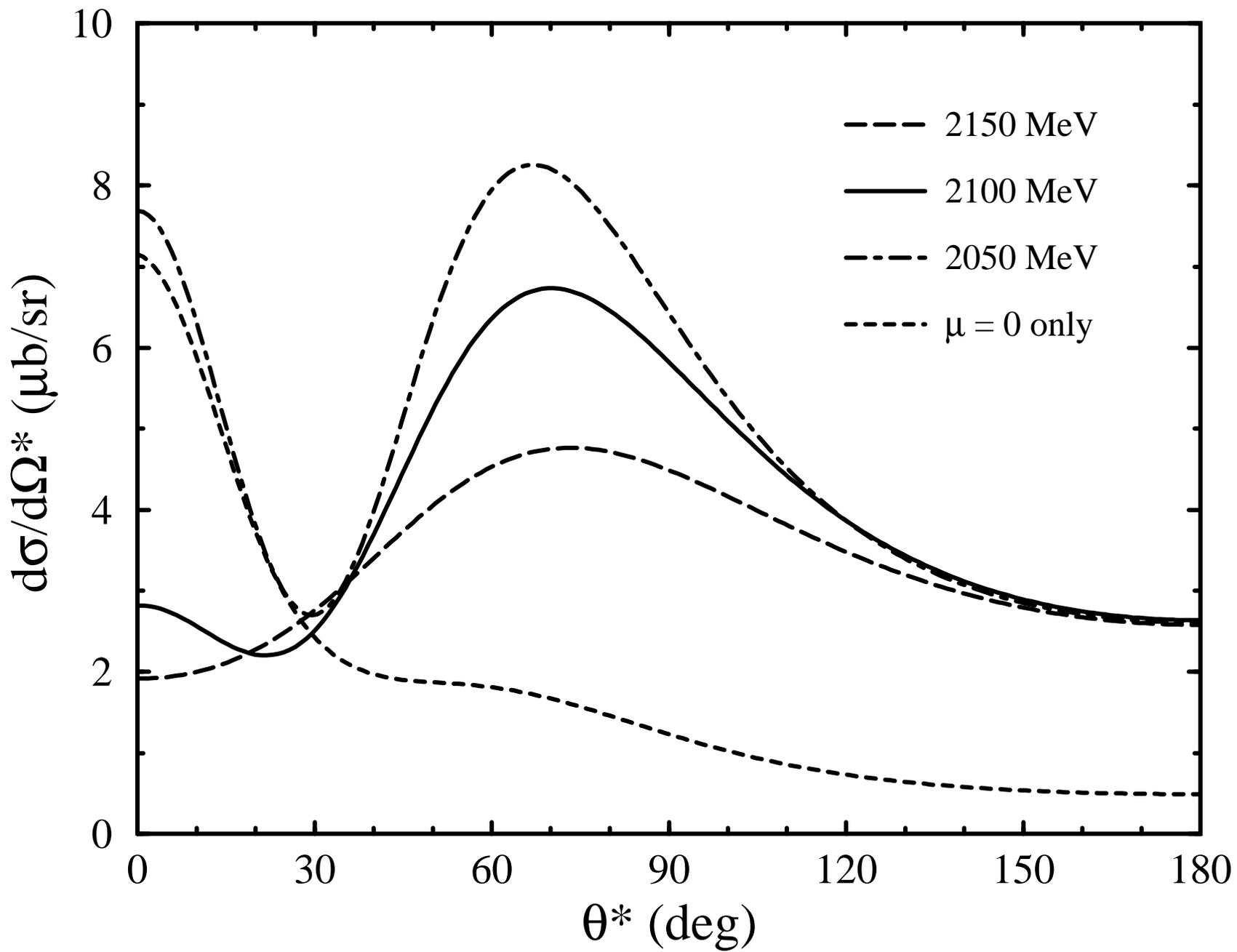
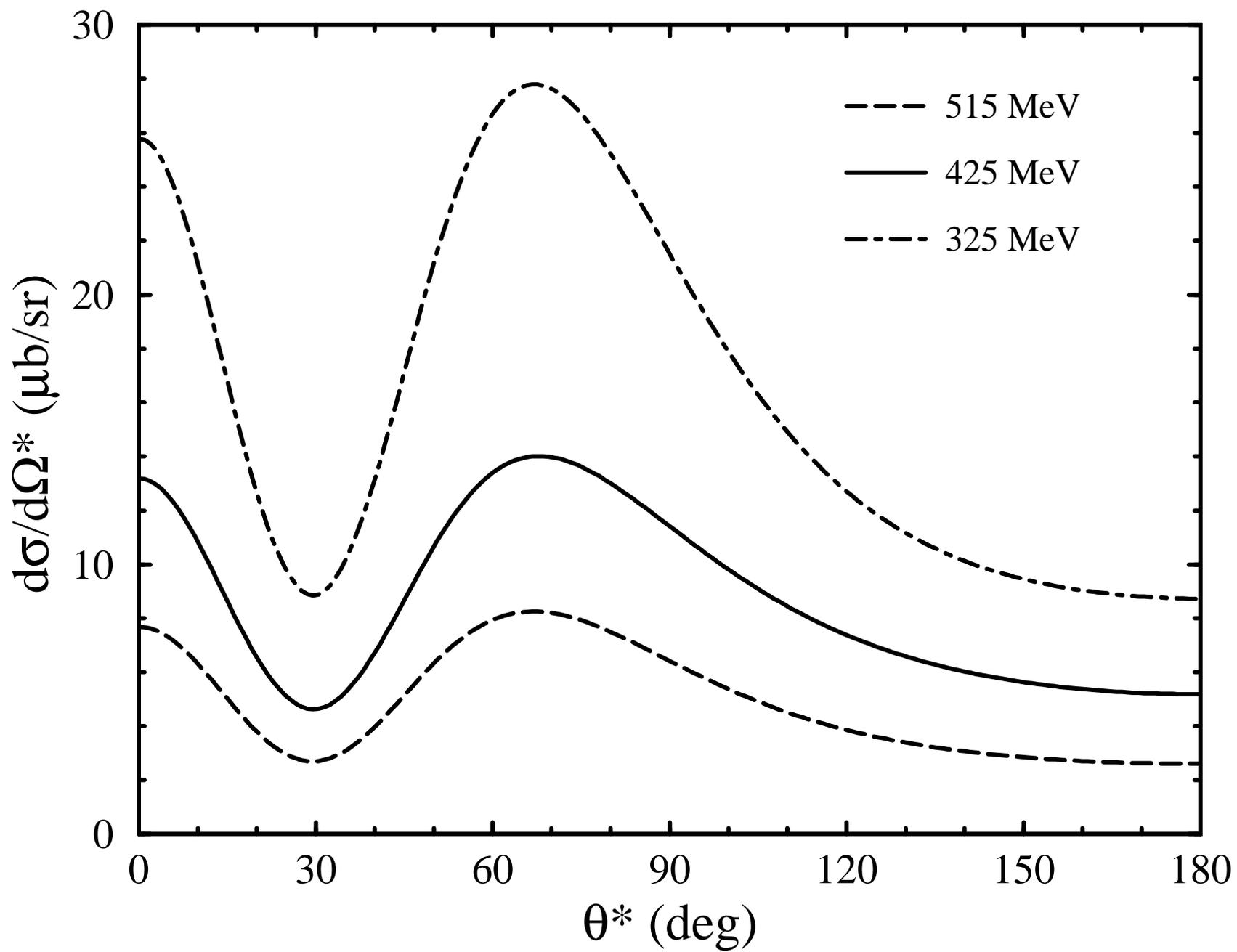
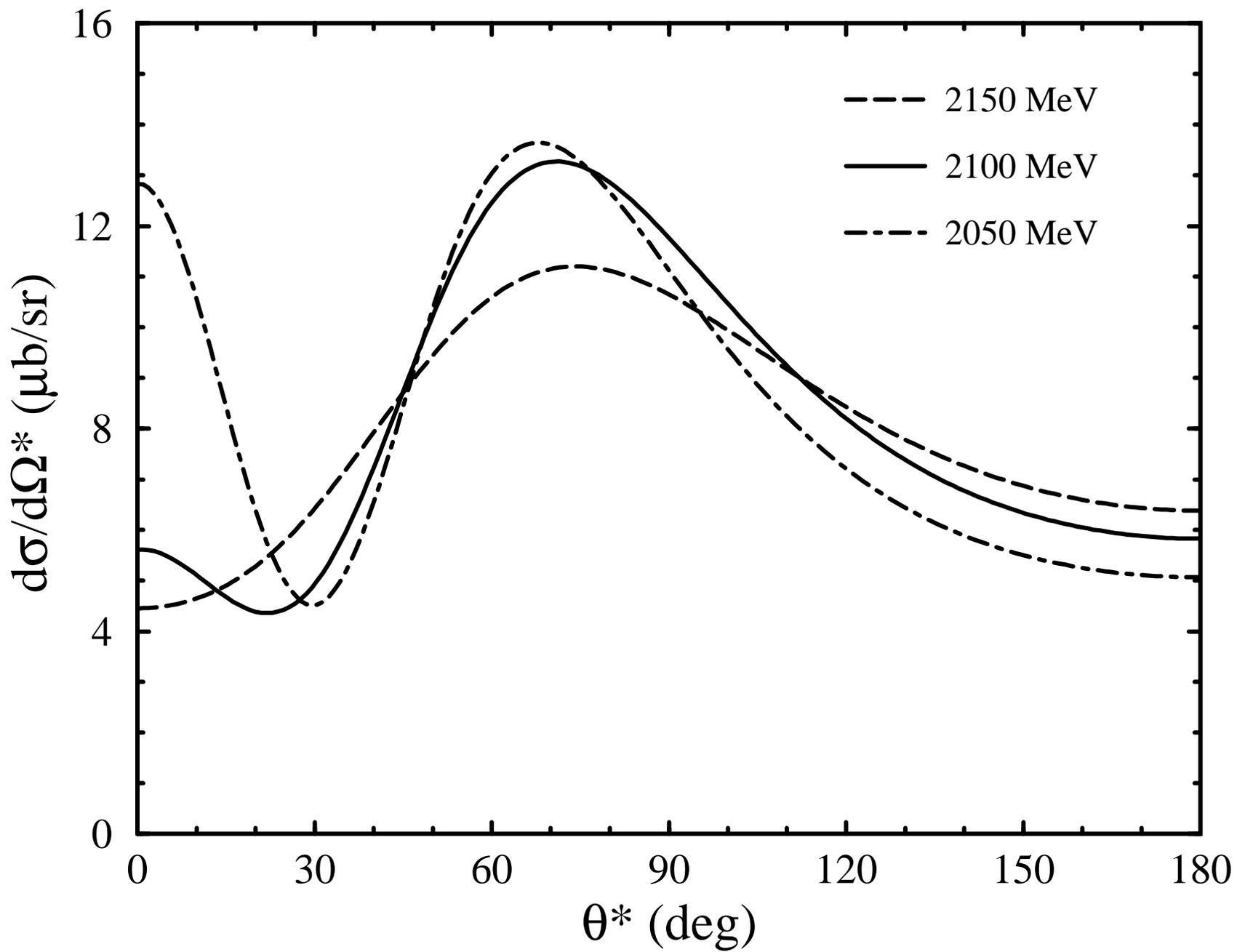
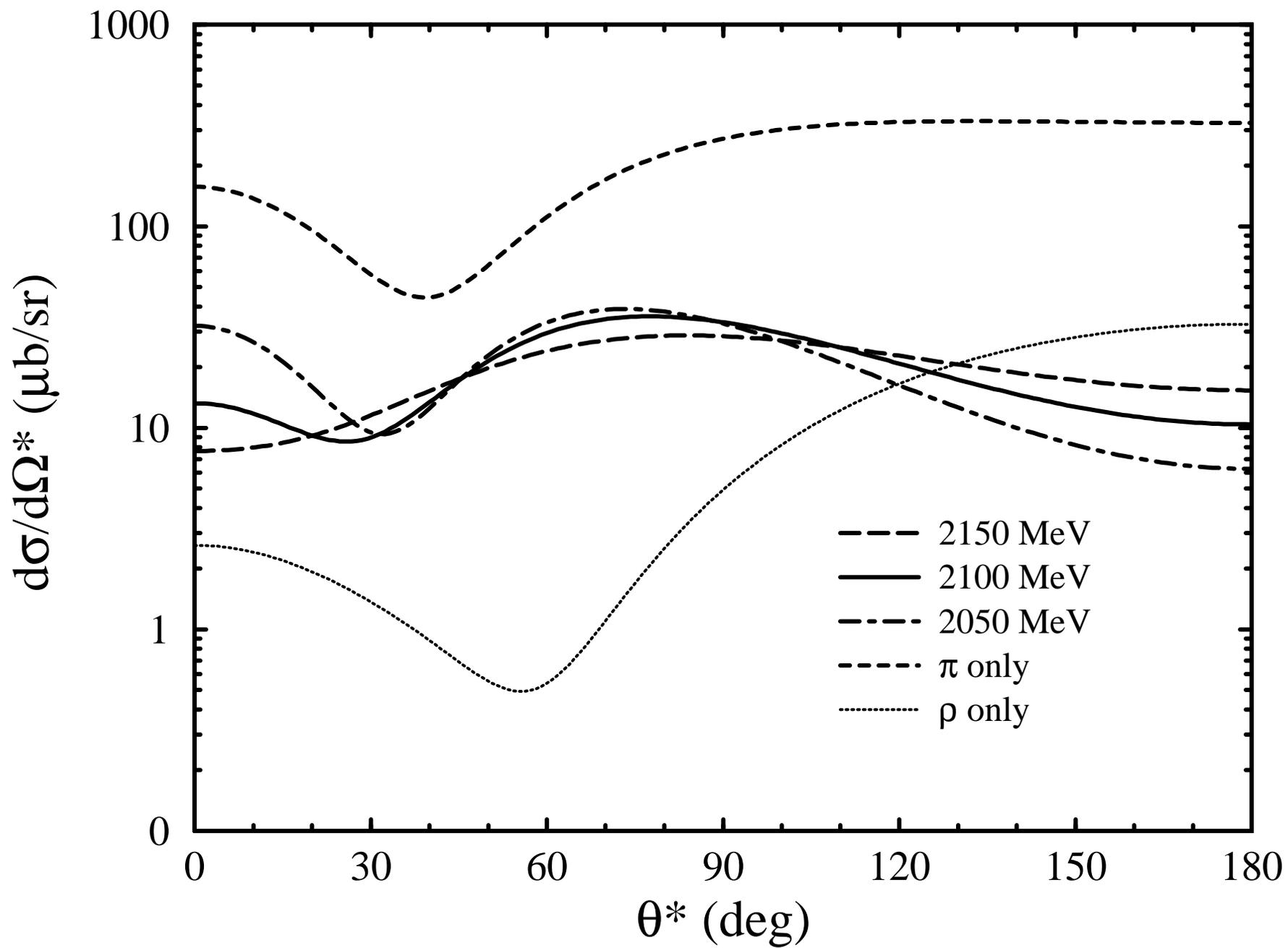
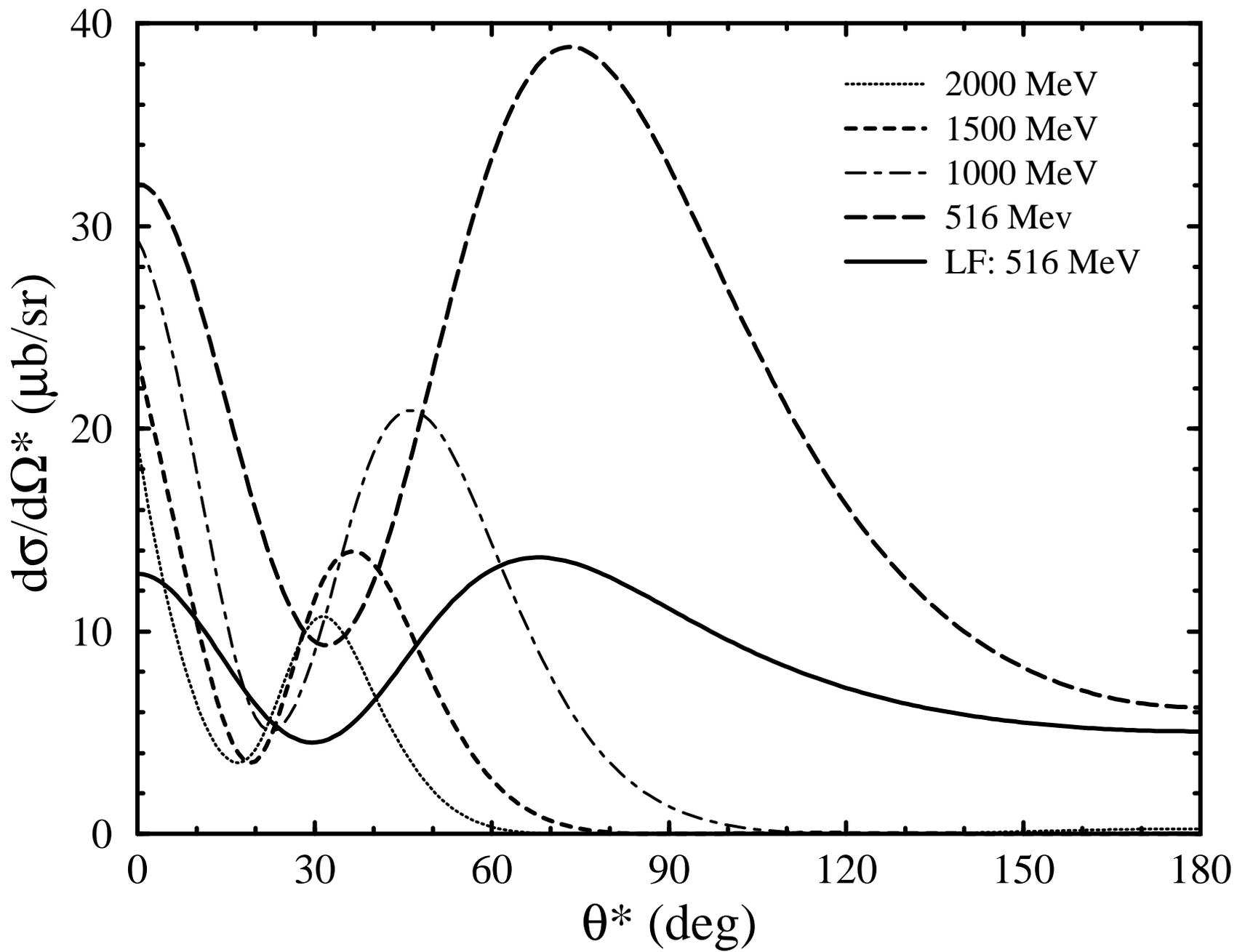
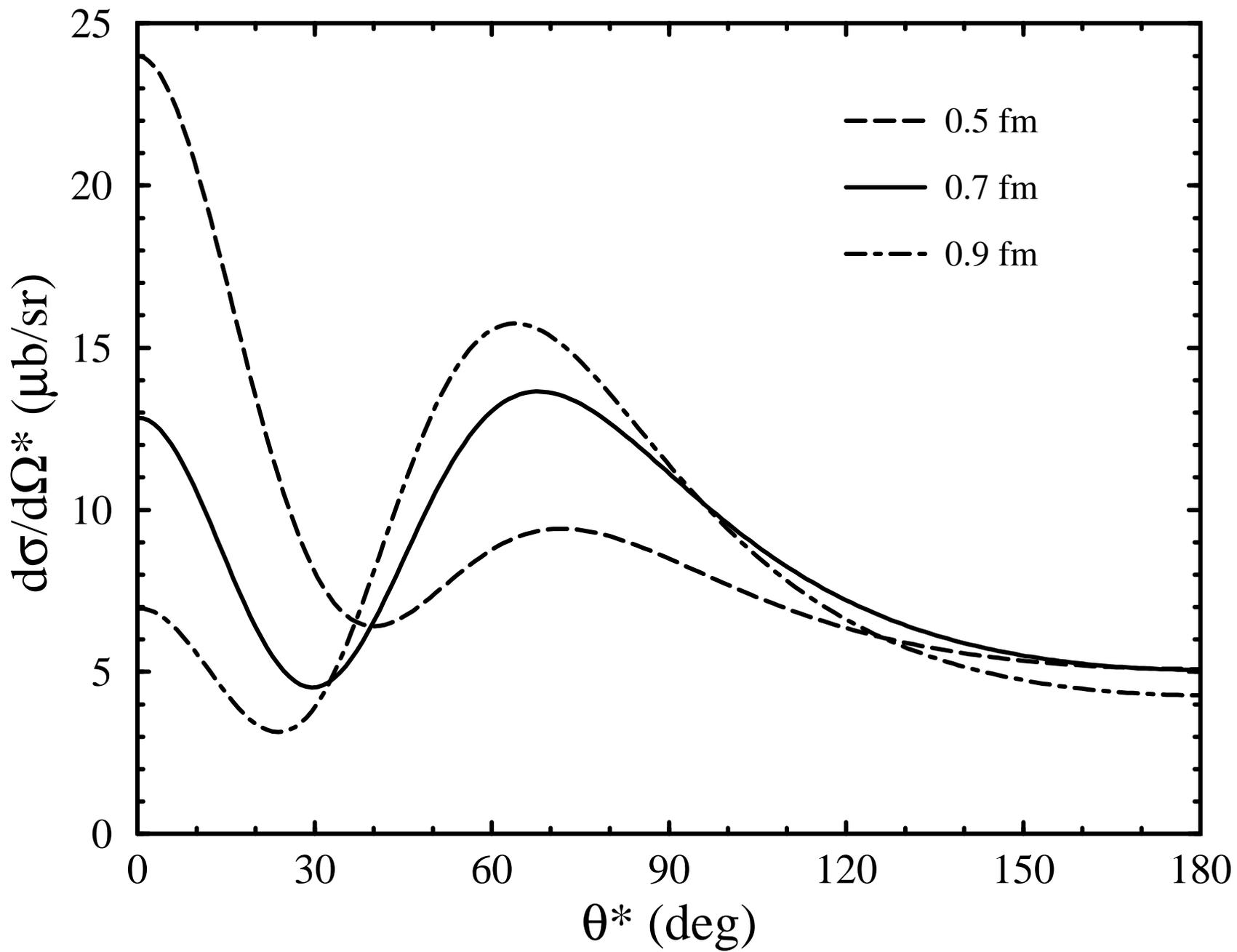
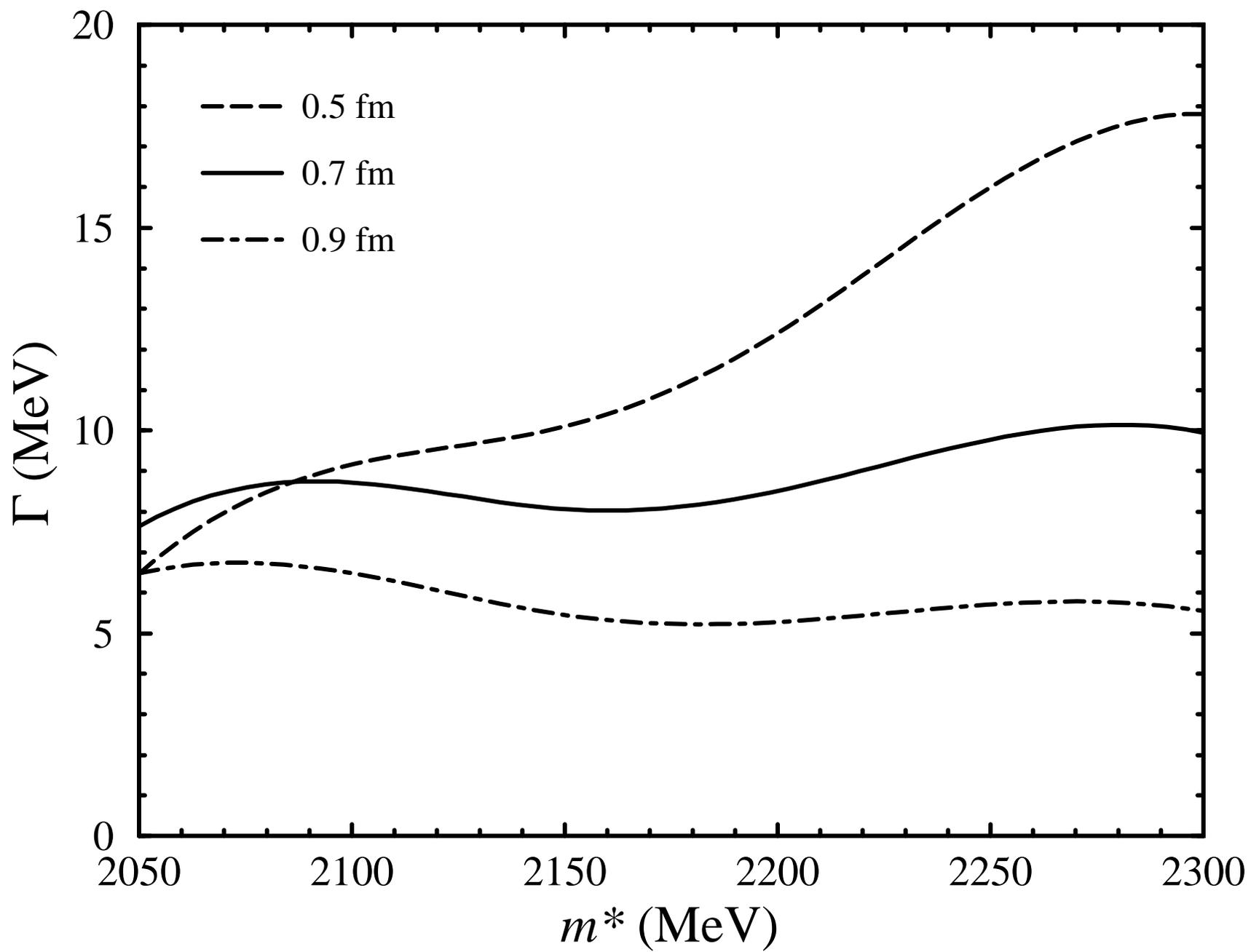
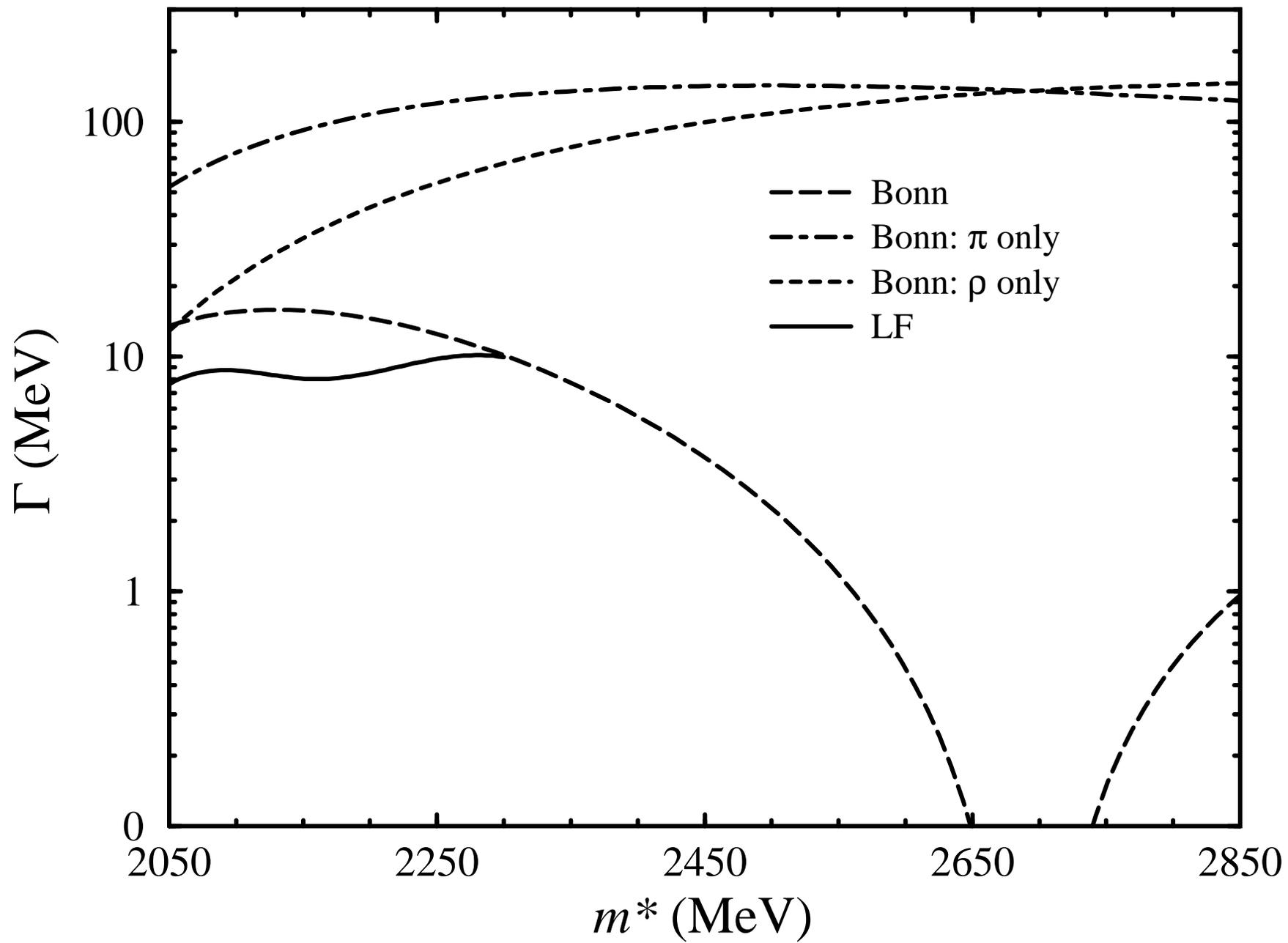
\end{document}